\documentclass[twocolumn,notitlepage]{aastex61}
\usepackage{amsmath,amstext}
\usepackage[T1]{fontenc}
\usepackage[figure,figure*]{hypcap}
\usepackage{subfigure}
\usepackage{xcolor}
\usepackage{soul}
\setstcolor{red}

\newcommand{\lephare}{\textsc{LePhare }}
\newcommand{\ppxf}{\texttt{pPXF }}
\newcommand{\sextractor}{\textsc{SExtractor }}
\newcommand{\galfit}{\textsc{GalFit }}

\shorttitle{Evidence for Dynamically Driven Formation of the GW170817 Neutron Star Binary}
\shortauthors{Palmese et al.}
\nopagebreak

\begin{document}

\title[Evidence for Dynamically Driven Formation of the GW170817 Neutron Star Binary]{Evidence for Dynamically Driven Formation of the GW170817 Neutron Star Binary in NGC 4993}

\author{A.~Palmese}\affiliation{Department of Physics \& Astronomy, University College London, Gower Street, London, WC1E 6BT, UK}
\author{W.~Hartley}\affiliation{Department of Physics \& Astronomy, University College London, Gower Street, London, WC1E 6BT, UK}
\author{F.~Tarsitano}\affiliation{Institute for Particle Physics and Astrophysics, ETH Zurich, Wolfgang-Pauli-Strasse 27, CH-8093 Zurich, Switzerland}
\author{C.~Conselice}\affiliation{University of Nottingham, School of Physics and Astronomy, Nottingham NG7 2RD, UK}
\author{O.~Lahav}\affiliation{Department of Physics \& Astronomy, University College London, Gower Street, London, WC1E 6BT, UK}
\author{S.~Allam}\affiliation{Fermi National Accelerator Laboratory, P. O. Box 500, Batavia, IL 60510, USA}
\author{J.~Annis}\affiliation{Fermi National Accelerator Laboratory, P. O. Box 500, Batavia, IL 60510, USA}
\author{H.~Lin}\affiliation{Fermi National Accelerator Laboratory, P. O. Box 500, Batavia, IL 60510, USA}
\author{M.~Soares-Santos}
\affiliation{Department of Physics, Brandeis University, Waltham, MA 02453, USA}
\affiliation{Fermi National Accelerator Laboratory, P. O. Box 500, Batavia, IL 60510, USA}
\author{D.~Tucker}\affiliation{Fermi National Accelerator Laboratory, P. O. Box 500, Batavia, IL 60510, USA} 
\author{D.~Brout}\affiliation{Department of Physics and Astronomy, University of Pennsylvania, Philadelphia, PA 19104, USA}
\author{M.~Banerji}\affiliation{Institute of Astronomy, University of Cambridge, Madingley Road, Cambridge CB3 0HA, UK}\affiliation{Kavli Institute for Cosmology, University of Cambridge, Madingley Road, Cambridge CB3 0HA, UK}
\author{K.~Bechtol}\affiliation{LSST, 933 North Cherry Avenue, Tucson, AZ 85721, USA}
\author{H.~T.~Diehl}\affiliation{Fermi National Accelerator Laboratory, P. O. Box 500, Batavia, IL 60510, USA}
\author{A.~Fruchter}\affiliation{Space Telescope Science Institute, 3700 San Martin Dr., Baltimore, MD 21218}
\author{J.~Garc\'ia-Bellido}\affiliation{Instituto de Fisica Teorica UAM/CSIC, Universidad Autonoma de Madrid, 28049 Madrid, Spain}
\author{K.~Herner}\affiliation{Fermi National Accelerator Laboratory, P. O. Box 500, Batavia, IL 60510, USA}
\author{A.~J.~Levan}\affiliation{Department of Physics, University of Warwick, Coventry, CV4 7AL, UK}
\author{T.~S.~Li} \affiliation{Fermi National Accelerator Laboratory, P. O. Box 500, Batavia, IL 60510, USA}
\author{C.~Lidman}\affiliation{ARC Centre of Excellence for All-sky Astrophysics (CAASTRO)}\affiliation{Australian Astronomical Observatory, North Ryde, NSW 2113, Australia}
\author{K.~Misra}\affiliation{Aryabhatta Research Institute of observational sciencES (ARIES), Manora Peak,  Nainital 263 001 India}
\author{M.~Sako}\affiliation{Department of Physics and Astronomy, University of Pennsylvania, Philadelphia, PA 19104, USA}
\author{D.~Scolnic}\affiliation{Kavli Institute for Cosmological Physics, University of Chicago, Chicago, IL 60637, USA}
\author{M.~Smith}\affiliation{School of Physics and Astronomy, University of Southampton,  Southampton, SO17 1BJ, UK}
 
\author{T.~M.~C.~Abbott}\affiliation{Cerro Tololo Inter-American Observatory, National Optical Astronomy Observatory, Casilla 603, La Serena, Chile}
\author{F.~B.~Abdalla}\affiliation{Department of Physics \& Astronomy, University College London, Gower Street, London, WC1E 6BT, UK}\affiliation{Department of Physics and Electronics, Rhodes University, PO Box 94, Grahamstown, 6140, South Africa}
\author{A.~Benoit-L{\'e}vy}\affiliation{CNRS, UMR 7095, Institut d'Astrophysique de Paris, F-75014, Paris, France}\affiliation{Department of Physics \& Astronomy, University College London, Gower Street, London, WC1E 6BT, UK}\affiliation{Sorbonne Universit\'es, UPMC Univ Paris 06, UMR 7095, Institut d'Astrophysique de Paris, F-75014, Paris, France}
\author{E.~Bertin}\affiliation{CNRS, UMR 7095, Institut d'Astrophysique de Paris, F-75014, Paris, France}
\affiliation{Sorbonne Universit\'es, UPMC Univ Paris 06, UMR 7095, Institut d'Astrophysique de Paris, F-75014, Paris, France}
\author{D.~Brooks}\affiliation{Department of Physics \& Astronomy, University College London, Gower Street, London, WC1E 6BT, UK}
\author{E.~Buckley-Geer}\affiliation{Fermi National Accelerator Laboratory, P. O. Box 500, Batavia, IL 60510, USA}
\author{A.~Carnero~Rosell}\affiliation{Laborat\'orio Interinstitucional de e-Astronomia - LIneA, Rua Gal. Jos\'e Cristino 77, Rio de Janeiro, RJ - 20921-400, Brazil}
\affiliation{Observat\'orio Nacional, Rua Gal. Jos\'e Cristino 77, Rio de Janeiro, RJ - 20921-400, Brazil}
\author{M.~Carrasco~Kind}\affiliation{Department of Astronomy, University of Illinois, 1002 W. Green Street, Urbana, IL 61801, USA}
\affiliation{National Center for Supercomputing Applications, 1205 West Clark St., Urbana, IL 61801, USA}
\author{J.~Carretero}\affiliation{Institut de F\'{\i}sica d'Altes Energies (IFAE), The Barcelona Institute of Science and Technology, Campus UAB, 08193 Bellaterra (Barcelona) Spain}
\author{F.~J.~Castander}\affiliation{Institute of Space Sciences, IEEC-CSIC, Campus UAB, Carrer de Can Magrans, s/n,  08193 Barcelona, Spain}
\author{C.~E.~Cunha}\affiliation{Kavli Institute for Particle Astrophysics \& Cosmology, P. O. Box 2450, Stanford University, Stanford, CA 94305, USA}
\author{C.~B.~D'Andrea}\affiliation{Department of Physics and Astronomy, University of Pennsylvania, Philadelphia, PA 19104, USA}
\author{L.~N.~da Costa}\affiliation{Laborat\'orio Interinstitucional de e-Astronomia - LIneA, Rua Gal. Jos\'e Cristino 77, Rio de Janeiro, RJ - 20921-400, Brazil}
\affiliation{Observat\'orio Nacional, Rua Gal. Jos\'e Cristino 77, Rio de Janeiro, RJ - 20921-400, Brazil}
\author{C.~Davis}\affiliation{Kavli Institute for Particle Astrophysics \& Cosmology, P. O. Box 2450, Stanford University, Stanford, CA 94305, USA}
\author{D.~L.~DePoy}\affiliation{George P. and Cynthia Woods Mitchell Institute for Fundamental Physics and Astronomy, and Department of Physics and Astronomy, Texas A\&M University, College Station, TX 77843,  USA}
\author{S.~Desai}\affiliation{Department of Physics, IIT Hyderabad, Kandi, Telangana 502285, India}
\author{J.~P.~Dietrich}\affiliation{Excellence Cluster Universe, Boltzmannstr.\ 2, 85748 Garching, Germany}
\affiliation{Faculty of Physics, Ludwig-Maximilians-Universit\"at, Scheinerstr. 1, 81679 Munich, Germany}
\author{P.~Doel}\affiliation{Department of Physics \& Astronomy, University College London, Gower Street, London, WC1E 6BT, UK}
\author{A.~Drlica-Wagner}\affiliation{Fermi National Accelerator Laboratory, P. O. Box 500, Batavia, IL 60510, USA}
\author{T.~F.~Eifler}\affiliation{Department of Physics, California Institute of Technology, Pasadena, CA 91125, USA}
\affiliation{Jet Propulsion Laboratory, California Institute of Technology, 4800 Oak Grove Dr., Pasadena, CA 91109, USA}
\author{A.~E.~Evrard}\affiliation{Department of Astronomy, University of Michigan, Ann Arbor, MI 48109, USA}
\affiliation{Department of Physics, University of Michigan, Ann Arbor, MI 48109, USA}
\author{B.~Flaugher}\affiliation{Fermi National Accelerator Laboratory, P. O. Box 500, Batavia, IL 60510, USA}
\author{P.~Fosalba}\affiliation{Institute of Space Sciences, IEEC-CSIC, Campus UAB, Carrer de Can Magrans, s/n,  08193 Barcelona, Spain}
\author{J.~Frieman}\affiliation{Fermi National Accelerator Laboratory, P. O. Box 500, Batavia, IL 60510, USA}
\affiliation{Kavli Institute for Cosmological Physics, University of Chicago, Chicago, IL 60637, USA}
\author{E.~Gaztanaga}\affiliation{Institute of Space Sciences, IEEC-CSIC, Campus UAB, Carrer de Can Magrans, s/n,  08193 Barcelona, Spain}
\author{D.~W.~Gerdes}\affiliation{Department of Astronomy, University of Michigan, Ann Arbor, MI 48109, USA}\affiliation{Department of Physics, University of Michigan, Ann Arbor, MI 48109, USA}
\author{T.~Giannantonio}\affiliation{Institute of Astronomy, University of Cambridge, Madingley Road, Cambridge CB3 0HA, UK}
\affiliation{Kavli Institute for Cosmology, University of Cambridge, Madingley Road, Cambridge CB3 0HA, UK}
\affiliation{Universit\"ats-Sternwarte, Fakult\"at f\"ur Physik, Ludwig-Maximilians Universit\"at M\"unchen, Scheinerstr. 1, 81679 M\"unchen, Germany}
\author{D.~Gruen}\affiliation{Kavli Institute for Particle Astrophysics \& Cosmology, P. O. Box 2450, Stanford University, Stanford, CA 94305, USA}
\affiliation{SLAC National Accelerator Laboratory, Menlo Park, CA 94025, USA}
\author{R.~A.~Gruendl}\affiliation{Department of Astronomy, University of Illinois, 1002 W. Green Street, Urbana, IL 61801, USA}
\affiliation{National Center for Supercomputing Applications, 1205 West Clark St., Urbana, IL 61801, USA}
\author{J.~Gschwend}\affiliation{Laborat\'orio Interinstitucional de e-Astronomia - LIneA, Rua Gal. Jos\'e Cristino 77, Rio de Janeiro, RJ - 20921-400, Brazil}
\affiliation{Observat\'orio Nacional, Rua Gal. Jos\'e Cristino 77, Rio de Janeiro, RJ - 20921-400, Brazil}
\author{G.~Gutierrez}\affiliation{Fermi National Accelerator Laboratory, P. O. Box 500, Batavia, IL 60510, USA}
\author{K.~Honscheid}\affiliation{Center for Cosmology and Astro-Particle Physics, The Ohio State University, Columbus, OH 43210, USA}
\affiliation{Department of Physics, The Ohio State University, Columbus, OH 43210, USA}
\author{B.~Jain}\affiliation{Department of Physics and Astronomy, University of Pennsylvania, Philadelphia, PA 19104, USA}
\author{D.~J.~James}\affiliation{Astronomy Department, University of Washington, Box 351580, Seattle, WA 98195, USA}
\author{T.~Jeltema}\affiliation{Santa Cruz Institute for Particle Physics, Santa Cruz, CA 95064, USA}
\author{M.~W.~G.~Johnson}\affiliation{National Center for Supercomputing Applications, 1205 West Clark St., Urbana, IL 61801, USA}
\author{M.~D.~Johnson}\affiliation{National Center for Supercomputing Applications, 1205 West Clark St., Urbana, IL 61801, USA}
\author{E.~Krause}\affiliation{Kavli Institute for Particle Astrophysics \& Cosmology, P. O. Box 2450, Stanford University, Stanford, CA 94305, USA}
\author{R.~Kron}\affiliation{Fermi National Accelerator Laboratory, P. O. Box 500, Batavia, IL 60510, USA}
\affiliation{Kavli Institute for Cosmological Physics, University of Chicago, Chicago, IL 60637, USA}
\author{K.~Kuehn}\affiliation{Australian Astronomical Observatory, North Ryde, NSW 2113, Australia}
\author{S.~Kuhlmann}\affiliation{Argonne National Laboratory, 9700 South Cass Avenue, Lemont, IL 60439, USA}
\author{N.~Kuropatkin}\affiliation{Fermi National Accelerator Laboratory, P. O. Box 500, Batavia, IL 60510, USA}
\author{M.~Lima}\affiliation{Departamento de F\'isica Matem\'atica, Instituto de F\'isica, Universidade de S\~ao Paulo, CP 66318, S\~ao Paulo, SP, 05314-970, Brazil}
\affiliation{Laborat\'orio Interinstitucional de e-Astronomia - LIneA, Rua Gal. Jos\'e Cristino 77, Rio de Janeiro, RJ - 20921-400, Brazil}
\author{M.~A.~G.~Maia}\affiliation{Laborat\'orio Interinstitucional de e-Astronomia - LIneA, Rua Gal. Jos\'e Cristino 77, Rio de Janeiro, RJ - 20921-400, Brazil}
\affiliation{Observat\'orio Nacional, Rua Gal. Jos\'e Cristino 77, Rio de Janeiro, RJ - 20921-400, Brazil}
\author{M.~March}\affiliation{Department of Physics and Astronomy, University of Pennsylvania, Philadelphia, PA 19104, USA}
\author{J.~L.~Marshall}\affiliation{George P. and Cynthia Woods Mitchell Institute for Fundamental Physics and Astronomy, and Department of Physics and Astronomy, Texas A\&M University, College Station, TX 77843,  USA}
\author{R.~G.~McMahon}\affiliation{Institute of Astronomy, University of Cambridge, Madingley Road, Cambridge CB3 0HA, UK}
\affiliation{Kavli Institute for Cosmology, University of Cambridge, Madingley Road, Cambridge CB3 0HA, UK}
\author{F.~Menanteau}\affiliation{Department of Astronomy, University of Illinois, 1002 W. Green Street, Urbana, IL 61801, USA}
\affiliation{National Center for Supercomputing Applications, 1205 West Clark St., Urbana, IL 61801, USA}
\author{C.~J.~Miller}\affiliation{Department of Astronomy, University of Michigan, Ann Arbor, MI 48109, USA}
\affiliation{Department of Physics, University of Michigan, Ann Arbor, MI 48109, USA}
\author{R.~Miquel}\affiliation{Instituci\'o Catalana de Recerca i Estudis Avan\c{c}ats, E-08010 Barcelona, Spain}
\affiliation{Institut de F\'{\i}sica d'Altes Energies (IFAE), The Barcelona Institute of Science and Technology, Campus UAB, 08193 Bellaterra (Barcelona) Spain}
\author{E.~Neilsen}\affiliation{Fermi National Accelerator Laboratory, P. O. Box 500, Batavia, IL 60510, USA}
\author{R.~L.~C.~Ogando}\affiliation{Laborat\'orio Interinstitucional de e-Astronomia - LIneA, Rua Gal. Jos\'e Cristino 77, Rio de Janeiro, RJ - 20921-400, Brazil}
\affiliation{Observat\'orio Nacional, Rua Gal. Jos\'e Cristino 77, Rio de Janeiro, RJ - 20921-400, Brazil}
\author{A.~A.~Plazas}\affiliation{Jet Propulsion Laboratory, California Institute of Technology, 4800 Oak Grove Dr., Pasadena, CA 91109, USA}
\author{K.~Reil}\affiliation{SLAC National Accelerator Laboratory, Menlo Park, CA 94025, USA}
\author{A.~K.~Romer}\affiliation{Department of Physics and Astronomy, Pevensey Building, University of Sussex, Brighton, BN1 9QH, UK}
\author{E.~Sanchez}\affiliation{Centro de Investigaciones Energ\'eticas, Medioambientales y Tecnol\'ogicas (CIEMAT), Madrid, Spain}
\author{R.~Schindler}\affiliation{SLAC National Accelerator Laboratory, Menlo Park, CA 94025, USA}
\author{R.~C.~Smith}\affiliation{Cerro Tololo Inter-American Observatory, National Optical Astronomy Observatory, Casilla 603, La Serena, Chile}
\author{F.~Sobreira}\affiliation{Instituto de F\'isica Gleb Wataghin, Universidade Estadual de Campinas, 13083-859, Campinas, SP, Brazil}
\affiliation{Laborat\'orio Interinstitucional de e-Astronomia - LIneA, Rua Gal. Jos\'e Cristino 77, Rio de Janeiro, RJ - 20921-400, Brazil}
\author{E.~Suchyta}\affiliation{Computer Science and Mathematics Division, Oak Ridge National Laboratory, Oak Ridge, TN 37831}
\author{M.~E.~C.~Swanson}\affiliation{National Center for Supercomputing Applications, 1205 West Clark St., Urbana, IL 61801, USA}
\author{G.~Tarle}\affiliation{Department of Physics, University of Michigan, Ann Arbor, MI 48109, USA}
\author{D.~Thomas}\affiliation{Institute of Cosmology \& Gravitation, University of Portsmouth, Portsmouth, PO1 3FX, UK}
\author{R.~C.~Thomas}\affiliation{Lawrence Berkeley National Laboratory, 1 Cyclotron Road, Berkeley, CA 94720, USA}
\author{A.~R.~Walker}\affiliation{Cerro Tololo Inter-American Observatory, National Optical Astronomy Observatory, Casilla 603, La Serena, Chile}
\author{J.~Weller}\affiliation{Excellence Cluster Universe, Boltzmannstr.\ 2, 85748 Garching, Germany}
\affiliation{Max Planck Institute for Extraterrestrial Physics, Giessenbachstrasse, 85748 Garching, Germany}
\affiliation{Universit\"ats-Sternwarte, Fakult\"at f\"ur Physik, Ludwig-Maximilians Universit\"at M\"unchen, Scheinerstr. 1, 81679 M\"unchen, Germany}
\author{Y.~Zhang}\affiliation{Fermi National Accelerator Laboratory, P. O. Box 500, Batavia, IL 60510, USA}
\author{J.~Zuntz}\affiliation{Institute for Astronomy, University of Edinburgh, Edinburgh EH9 3HJ, UK}

\begin{abstract}
\noindent We present a study of NGC 4993, the host galaxy of the GW170817 gravitational wave event, the GRB170817A short gamma--ray burst (sGRB) and the AT2017gfo kilonova. We use Dark Energy Camera imaging, AAT spectra and publicly available data, relating our findings to binary neutron star (BNS) formation scenarios and merger delay timescales. NGC4993 is a nearby early--type galaxy, with $i$-band S\'ersic index $n=4.0$ and low asymmetry ($A=0.04\pm 0.01$). These properties are unusual for sGRB hosts. However, NGC4993 presents shell--like structures and dust lanes indicative of a recent galaxy merger, with the optical transient located close to a shell. We constrain the star formation history (SFH) of the galaxy assuming that the galaxy merger produced a star formation burst, but find little to no on--going star formation in either spatially--resolved broadband SED or spectral fitting. We use the best--fit SFH to estimate the BNS merger rate in this type of galaxy, as $R_{NSM}^{gal}= 5.7^{+0.57}_{-3.3} \times 10^{-6} {\rm yr}^{-1}$. If star formation is the only considered BNS formation scenario, the expected number of BNS mergers from early--type galaxies detectable with LIGO during its first two observing seasons is $0.038^{+0.004}_{-0.022}$, as opposed to $\sim 0.5$ from all galaxy types. Hypothesizing that the binary formed due to dynamical interactions during the galaxy merger, the subsequent time elapsed can constrain the delay time of the BNS coalescence. By using velocity dispersion estimates and the position of the shells, we find that the galaxy merger occurred $t_{\rm mer}\lesssim 200~{\rm Myr}$ prior to the BNS coalescence. 

\end{abstract}


\section{Introduction}
The first identification of the optical counterpart \citep{MMApaper} of a gravitational wave (GW) signal \citep{ligobns} marks the beginning of a new era for multi-messenger astronomy. The coalescence of neutron stars is expected
to have strong optical and near-infrared signatures in the form of a
kilonova, the ejecta from which are heated by the decay of heavy nuclei produced via rapid neutron--capture processes ($r$-processes). Short Gamma--Ray Bursts (sGRB) are likely to be related to the same coalescence events (\citealt{sgrb}), but the formation of the binary and the physics involved in merging are still a matter of debate (\citealt{bns}; \citealt{bns2}).

\begin{figure*}
\centering
\includegraphics[width=1\textwidth]{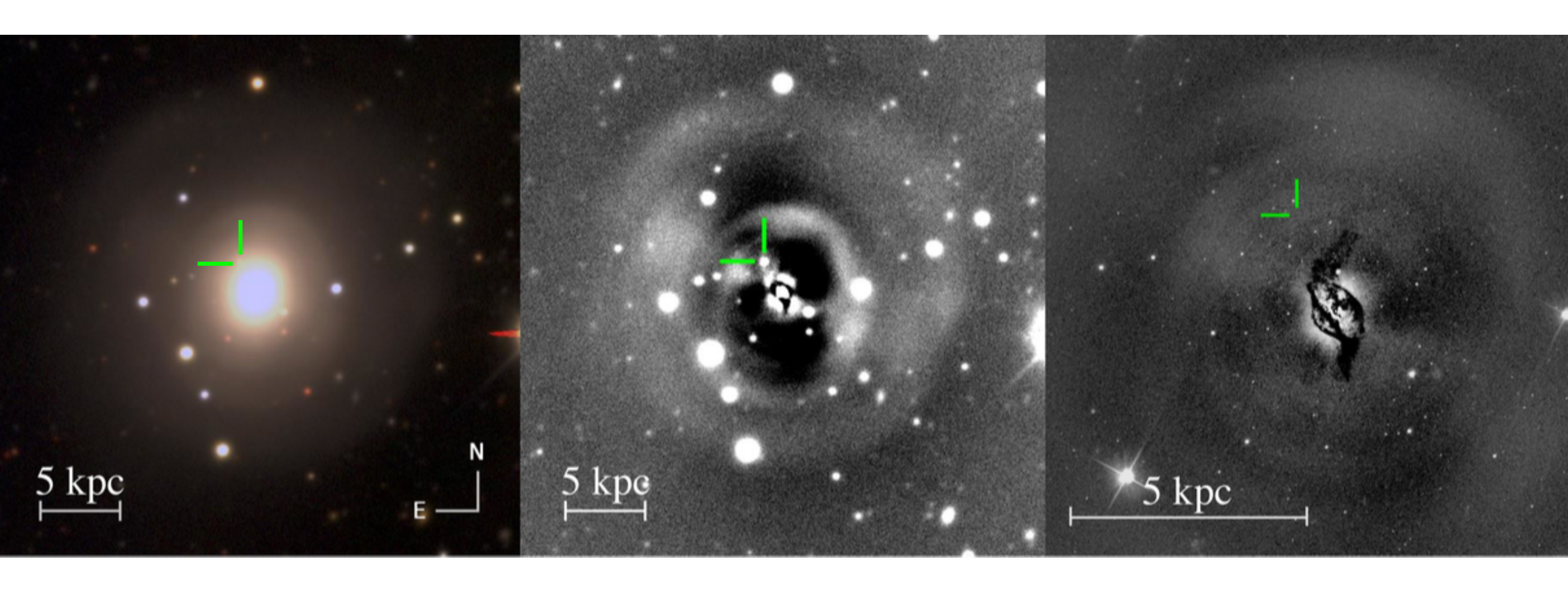}
\caption{\emph{Left panel:} DECam coadded image of NCG4993 in $gri$. Shell structures indicative of a recent galaxy merger are clearly visible. \emph{Middle panel:} $r$-band residuals from \galfit after subtraction of the best--fitting single S\'ersic light profile. \emph{Right panel:} F606W-band \textit{HST} ACS image with a 3 component galaxy model subtracted. Dust lanes crossing the centre of the galaxy are evident. The green lines show the position of the transient. The BNS counterpart is only present in the middle panel.
}\label{coaddimage}\end{figure*}

The optical counterpart to the binary neutron star (BNS) coalescence signal GW170817 was discovered independently by several collaborations using optical telescopes, including the Dark Energy Camera (DECam; \citealt{flaugher}) GW team \citep{marcelle17}. 
In this work we use this DECam data and supplement it with Hubble Space Telescope, Anglo--Australian Telescope (AAT) spectroscopic data and with publicly available datasets to understand the source in the context of its host galaxy and the local environment. 

In particular, we relate the BNS formation to the dynamics and stellar evolution of the host over time, asking whether the binary system was born as such, or whether dynamical interactions caused its formation. Dynamically--driven binary formation has been proposed for binary black holes (e.g. \citealt{bbh1}).

Previous studies (\citealt{1988MNRAS.235..813C}) classified this galaxy as an atypical elliptical galaxy with faint concentric shells and spectral features suggesting that the galaxy has undergone a merging event. Shells are visible as arcs of enhanced surface brightness corresponding to higher stellar densities around a galaxy center, and they are thought to be the relics of the infalling stars and interstellar matter from a galaxy merger. Several analytical and numerical studies support the galaxy merger scenario for the formation of shells in galaxies (e.g. \citealt{quinn}; \citealt{pop}), and show that the distribution of shells can constrain the time of the merger event. We study the evolution of this galaxy to
discern between different BNS formation scenarios and estimate the rate of BNS formation in early-type galaxies, using Dark Energy Survey (DES) data to place NGC4993 in the context of the galaxy population.

\section{data}\label{datasec}

\subsection{Photometric data: DECam, VHS and \textit{HST}}

The DECam images used in this work were taken as part of the DECam-GW follow up program between the nights of 2017 August 17 and September 1, using $ugrizY$ filters. We also use public $ugrizY$ DECam data from June 2015 to avoid contamination in the transient region. In addition, we extract $YJK$ data from the VISTA Hemisphere Survey (VHS; \citealt{vista}), covering the host galaxy.
The images are coadded and registered to a common pixel scale ($0.2636''$) using \texttt{SWARP} \citep{swarpref} with $3.5$ sigma clipping to remove cosmic ray artifacts. An RGB coadded image of the galaxy is presented in Figure \ref{coaddimage}. We build a $\chi^2$ detection image from the $r$, $i$ and $z$-band data and run \sextractor \citep{sextractor} in dual mode on the coadded images without performing template subtraction.

The photometry is corrected for galactic extinction. 
In order to compare the galaxy properties to a broader sample, we also use DES data from the first year of observations (Y1; \citealt{firstyear}).
We use \texttt{MAG\_AUTO} magnitudes unless otherwise stated.

NGC4993 was also observed during Hubble Space Telescope (\textit{HST}) Cycle 24
(PropID 14840, PI: Bellini) using ACS in F606W. The data were publicly released in April 2017 and were accessed via the Hubble Source Catalog (HSC;  \citealt{whitemore}).

\subsection{Spectroscopic data: 6dF and AAT}

The 6dF Galaxy Survey (\citealt{6dfgs}) final release (\citealt{6df}) includes an optical spectrum of the center of NGC4993 with an estimated redshift ($z=0.009680 \pm 0.000150$). 

Spectra of 14 galaxies with $v_{helio}\sim3000~{\rm km}~{\rm s}^{-1}$ and within one degree radius of NGC4993 were obtained in one target of opportunity exposure of the AAOmega spectrograph at the Anglo--Australian Telescope (AAT) on 2017 August 27. Of those, 10 spectral fits passed quality cuts.
All the spectra used here are centered on their galaxy nucleus with a $2''$ aperture.


\section{Host morphology}\label{morphsec}

\subsection{CAS and Galfit}\label{paramfitting}

We begin our study of NGC4993 with an analysis of its morphological properties, employing the $CAS$ non-parameteric light quantification \citep{conselice} and parametric S\'ersic light profile fitting with \textsc{GalFit} \citep{Peng}. Both methods utilise a mask to exclude other sources in the image and the location of the kilonova event. The $CAS$ system is able to pick out the salient features of galaxy morphology, allowing galaxy types to be assigned and identifying objects that are likely to have undergone a recent major merger (see \citealt{conselice} for details). Meanwhile, fitting the light profile additionally provides us with an alternative estimate of the total magnitude and can reveal more subtle aspects of galaxy morphology within the residuals of the model-subtracted image.

\textsc{GalFit} is run on the DES and VHS images in two ways: band-by-band and simultaneously across all bands using a modified version, \textsc{GalFit-m} \citep{galfitm}. In the second case the S\'ersic fitting parameters are allowed to vary with wavelength as a second-order polynomial. We extract the PSF model required by \textsc{GalFit} from the coadd images with \textsc{PSFEx} \citep{psfex} and initialize the fitting parameters based on measurements of the galaxy from \textsc{sextractor}. All parameters are left free without constraints, except for the central position in the single-band fits. This is allowed to vary by only $\pm 1$ pixel as it is well-constrained by \sextractor already.

In order to assess the stability of \textsc{GalFit} and obtain an estimate of the uncertainties on the measurements, each single--band run is performed 10,560 times, varying the inputs around their nominal values. We take the median as our final measurement and the standard deviation as the uncertainty.

\begin{figure}
\centering
\includegraphics[width=0.5\textwidth]{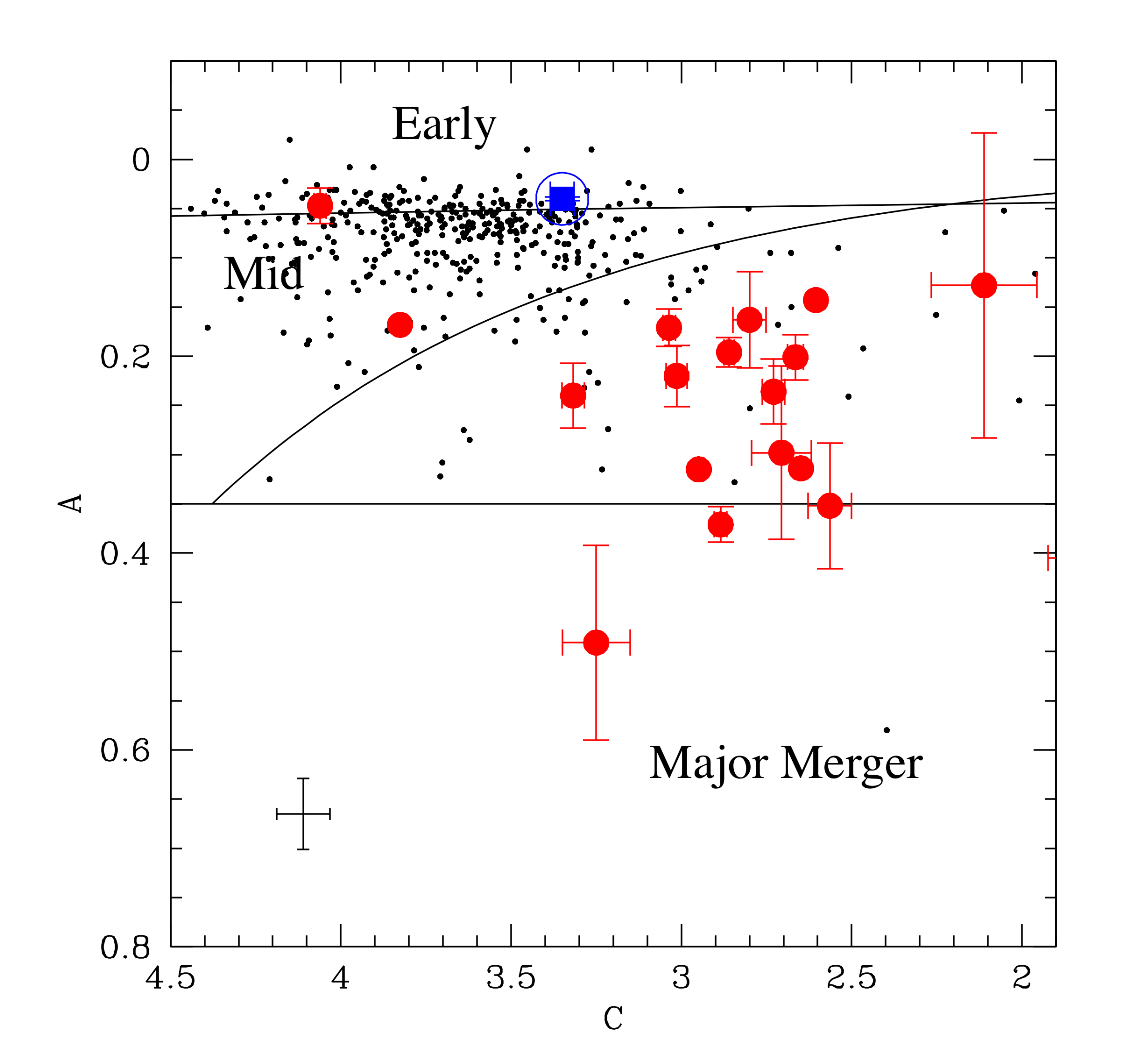}\caption{Concentration versus asymmetry for NGC4993 (in blue), compared to a sGRB hosts sample (Conseliece et al., in prep.) in red, and field galaxies (black dots) with stellar mass within $\pm 0.2$ dex of NGC4993 value and redshift $z<0.2$. The lines separate different Hubble types as shown in \citet{conselice}.}\label{sgrb}\end{figure}

\begin{table*}
\centering
\begin{tabular}{c|cccccc}
\hline
\hline

Filter & \texttt{MAG\_AUTO} & Mag & $r_e$ & $n$ & $\epsilon$ & $\theta$\\
\hline
$u$ & 14.24 & 14.15 & 61.8 & 3.2 & 0.15 & -13.9\\
$g$ & 12.95 & 12.80 & 62.5 & 3.4 & 0.15 & -12.8\\
$r$ & 12.08 & 11.90 & 63.5 & 3.7 & 0.16 & -11.2\\
$i$ & 11.65 & 11.45 & 64.4 & 4.0 & 0.16 & -9.9\\
$z$ & 11.34 & 11.13 & 65.3 & 4.3 & 0.16 & -8.4\\
$Y$ & 11.13 & 10.96 & 65.7 & 4.4 & 0.16 & -7.7\\
$Y_{{\rm VHS}}$ & 11.27 & 11.00 & 65.9 & 4.5 & 0.16 & -7.5\\
$J$ & 11.00 & 10.77 & 67.3 & 5.0 & 0.17 & -5.2\\
$K_s$ & 11.08 & 10.68 & 72.9 & 6.7 & 0.19 & +3.5\\
\hline
 & & $\pm 5\times10^{-4}$ & $\pm 0.07$ & $\pm 3\times10^{-3}$ & $\pm 4\times10^{-5}$ & $\pm 5\times10^{-3}$ \\
\hline

\end{tabular}\caption{Outputs from \textsc{Galfit} parametric S\'ersic fits performed on the $ugrizY$ DECam coadd images and $YJKs$ VHS data. The fit was joint across bands, allowing the effective radius, $r_e$ (in pixels), S\'ersic index, $n$, ellipticity, $\epsilon=1-b/a$, and position angle, $\theta$, to vary with wavelength. One pixel corresponds to $0.2636''$. The final row lists indicative errors based on the single band analysis.}\label{tablefit}
\end{table*}

\subsection{Results}
\label{morphresults}
Following the definitions given in \citet{conselice}, we find: concentration $C=3.348\pm 0.035$, asymmetry $A=0.04\pm 0.01$, and clumpiness $S=0.05\pm 0.05$. These values are typical for an early-type galaxy. In Figure \ref{sgrb} we compare these values to field galaxies of similar masses (within 0.2 dex of NGC4993) and redshifts ($z<0.2$) from the GAMA survey, and to a sample of sGRB hosts (Conselice et al. in prep.) taken in F814W imaging from \textit{HST}. NGC4993 stands out as peculiar with respect to other GRB hosts: such objects tend to lie on the more highly asymmetric side of late-type galaxies.

The results from the single S\'ersic fit across all bands are summarized in Table \ref{tablefit} (the band-by-band fits give broadly consistent results). We find an increase in S\'ersic index towards redder bands and a rotation in the position angle. 
This rotation of bluer versus redder bands suggests there could be two superimposed stellar populations with differing orientations. This may have arisen during the course of the galaxy's secular evolution but could also be caused by a minor galaxy merger, as indicated by the presence of shells.

The middle panel of Figure \ref{coaddimage} shows DECam $r$-band residuals from \galfit and the position of the transient. At least four shell structures are clearly visible. The surface brightness radial profile from the residual image shows an excess at the shell positions of $\sim 25~{\rm mag ~arcsec^{-2}}$. Closer inspection with \textit{HST} data (right panel in Figure \ref{coaddimage}) reveals a possible further broad inner shell, on which the transient seems to lie, and obvious dust lanes (visible also as a negative residual in the DECam version).  In summary, these results provide compelling evidence for a recent minor galaxy merger in NCG4993, and the location of the kilonova event with respect to the shells leads us naturally to ask whether there is a causal connection between the two, for instance via dynamical interaction.  

The $r$-band absolute magnitude from a $4~{\rm sq.~arcsec}$ region around the transient location in the galaxy-subtracted template image is $-10.65$. This luminosity implies a rather high stellar density in the locale of the BNS coalescence, implying that dynamical interactions between stars are more probable in this region compared to typical galaxy stellar densities.

From Figure \ref{sgrb} we see that clear major galaxy mergers are unusual amongst sGRB hosts. Furthermore, the other sGRBs are at cosmological distances and thus are mostly undergoing extensive galaxy formation through star formation or merging.  If the hosts have to be related by some common features, this is an indication that NGC4993 has undergone some merging activity, but a minor merger such that the bulk morphology is still elliptical. We thus explore the possibility that the kilonova was a result of a recent galaxy merger in NGC4993.


\section{Photometric and spectroscopic SED}\label{sedsec}

\begin{figure*}
\hspace{-0.5cm}
\includegraphics[width=0.63\textwidth]{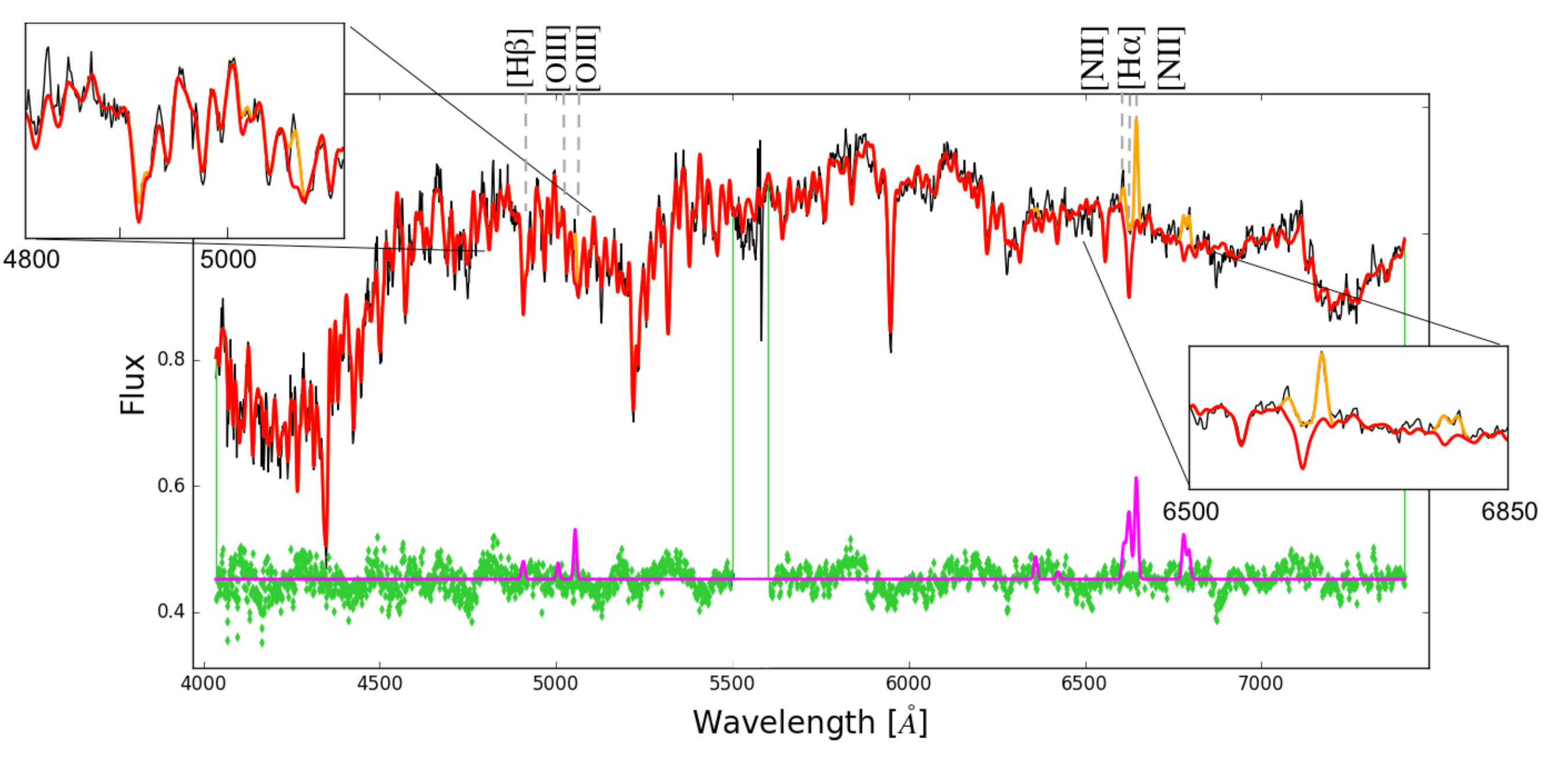}\hspace{-0.1cm}\vspace{-0.1cm}\includegraphics[width=0.4\textwidth]{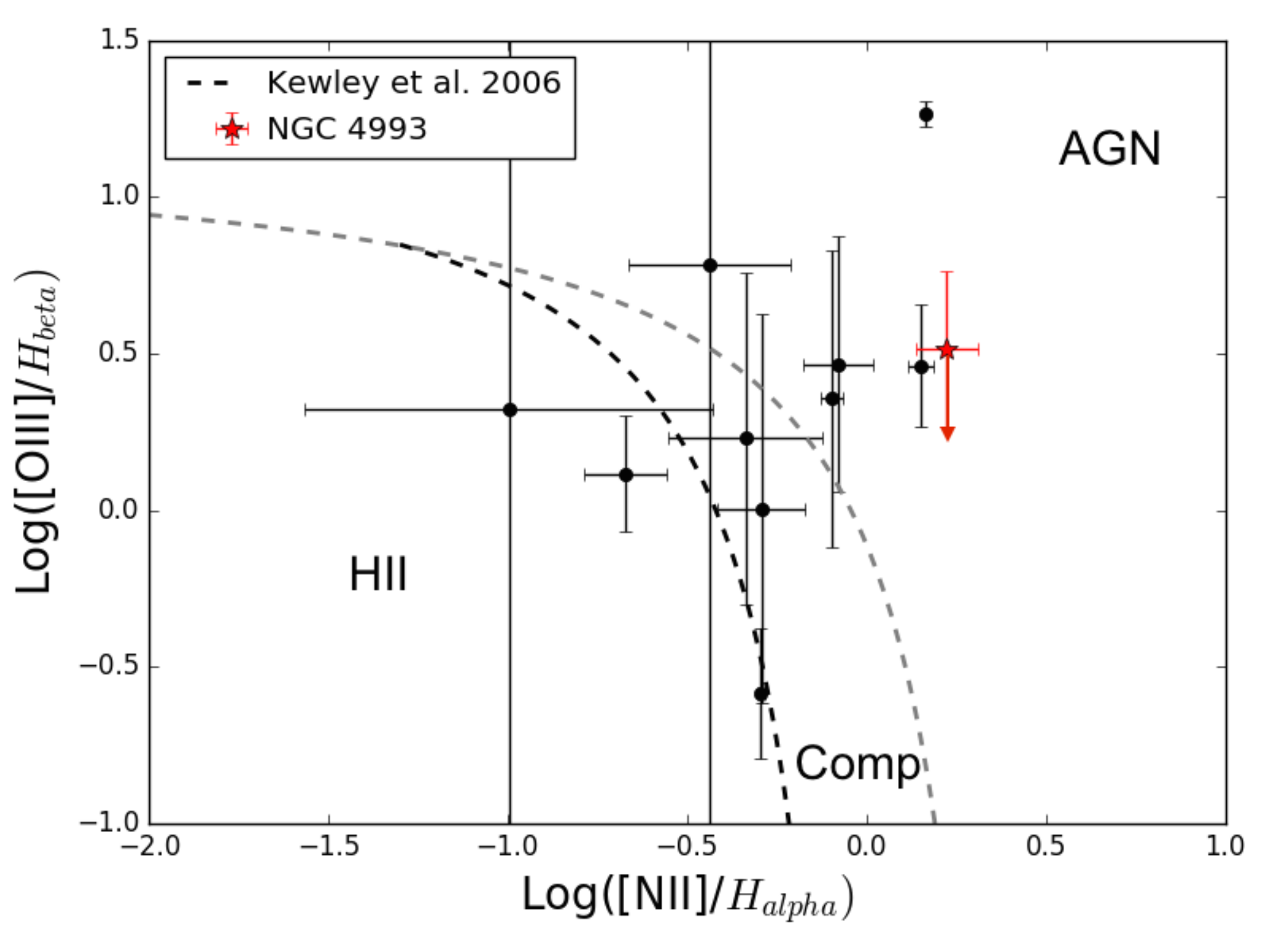}
\caption{\emph{Left panel:} spectroscopic fit of the 6dF optical spectrum. The black line is the observed spectrum,the red line is the \textsc{pPXF} fit for the stellar component, and the orange line is the best--fit including ionized gas emission lines. The zoomed panels show $H_\beta$, OIII, NII and $H_\alpha$ lines. The green points at the bottom are the fit residuals, while the purple line is the gas-only best--fit model spectrum. \emph{Right panel:} BPT diagram for NGC4993 (red star) and the other galaxies (black points) in the galaxy group with AAT spectra available. The dashed lines represent the \citet{2006MNRAS.372..961K} classification method for AGN, star--forming (HII) and composite (Comp) galaxies. Many of the group galaxies have very weak AGN or LINER-like emission. Error bars represent 1$\sigma$ error from the propagation of fit errors on line strengths.}\label{spec}\end{figure*}

\subsection{SED fitting methods}
We use \ppxf (\citealt{ppxf}; \citealt{ppxf1}), for the spectral fitting. It enables extraction of the stellar kinematics and stellar population from absorption line spectra of galaxies, using a maximum penalized likelihood approach. We use the Miles stellar libraries, and fit over the wavelengths $4000-7409 $ \AA, excluding the range $5500-5600$ \AA\, of the 6dF spectrum, where a strong sky line contaminates the flux.

We use \lephare (\citealt{arnouts}, \citealt{ilbertlephare}) for the broadband Spectral Energy Distribution (SED) fitting. 
We add a 0.05 systematic uncertainty in quadrature to the magnitudes. 
The simple stellar population (SSP) templates used are \citet{bc03}, with two metallicities ($Z_\odot$ and $2.5Z_\odot$), a \citet{chabrier} Initial Mass Function (IMF) and a Milky Way \citep{allen} extinction law. The SFH chosen is lognormal:
\begin{equation}
  \Psi(t,t_0,\tau)=\frac{1}{t\sqrt{2\pi\tau^2}}e^{-\frac{(\ln t-\ln t_0)^2}{2\tau^2}}\,,  \label{sfr} 
\end{equation}
as it is the most representative family of models with only two parameters \citep{lognorm}. Here $t_0$ and $\tau$ are the half--mass--time and width.

Motivated by our morphological analysis, we allow for an additional burst of recent SF. This is modeled as a Gaussian centered at $t_{burst}$ with width of 10 Myr and peaking at a fraction $0.4-0.1$ of the peak of the log-normal SFH (as no evidence for strong late SF is found).

The same templates are used to perform spatially--resolved SED fitting across DES+VHS coadded images within $10\times10$ pixels, including the galaxy dust extinction. The other sources in the field are masked out using the segmentation map output by \textsc{sextractor}.

\begin{figure}
\hspace{0.2cm}\includegraphics[width=0.45\textwidth]{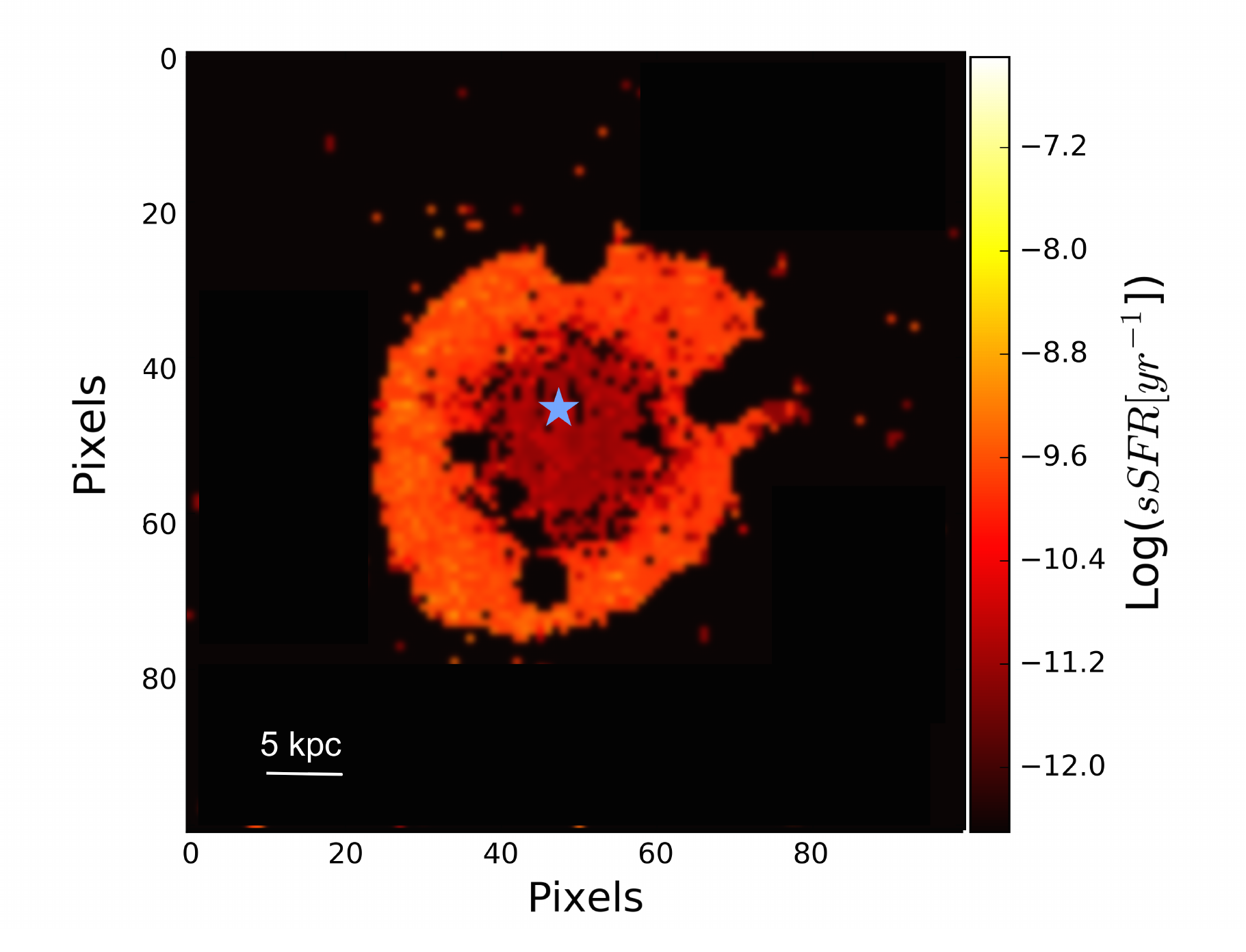}
\includegraphics[width=0.41\textwidth]{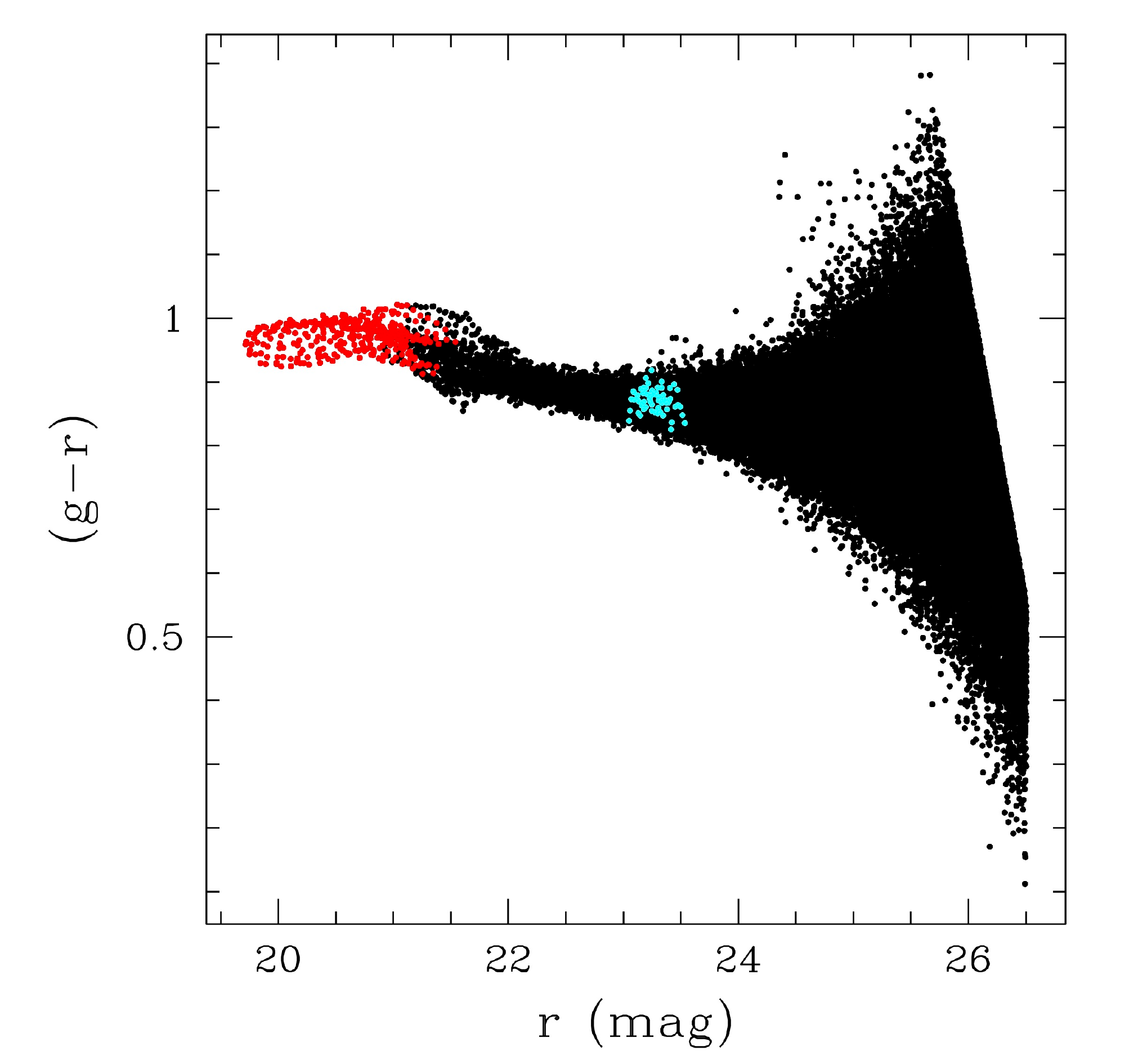}
\caption{\emph{Upper panel:} sSFR map resulting from the pixel SED fitting of DES+VHS bands. Other objects have been masked out. One pixel corresponds to $10\times 10$ DECam pixels, and to a physical size of 0.526 kpc at the galaxy redshift. 
\emph{Lower panel:} pixel color-magnitude diagram for the pixels covering NGC4993 from DECam $g$ and $r$ single epoch exposures taken previously to the BNS event. The core of the galaxy is shown in red, while the cyan points represent the $1.5''$ around the location of the BNS event ($10.6''$ from the center).}\label{smass_map}\end{figure}

\subsection{SED fitting results}\label{resultssec}
Figure \ref{spec} shows the best fit model of the 6dF spectrum, which results in a reduced $\chi^2$ of 1.22. An analysis of the mass fraction in age shows that part of the core galaxy stellar population has a supersolar metallicity, but the weighted mean value $\langle [M/H]\rangle=-0.012\pm 0.010$ is marginally consistent with solar metallicity. The mean age is $11.298\pm 0.054$ Gyr, and the mass-to-light ratio is $5.23\pm 0.15$ in $r$-band.

The stellar model fit reveals the existence of weak ionized gas emission lines. However, the line ratios from the fit suggests they are produced by a harder ionizing source than star-formation, formally lying in the AGN region of the Baldwin, Phillips \& Telervich (BPT; \citealt{BPTref}) diagram. \citet{blanchard} argue that there is a weak AGN present in the core of the galaxy on the basis of radio and X-ray emission, and so we conclude there to be no evidence of recent star-formation from the 6dF spectrum, irrespective of the highly uncertain ${\rm O\,\textsc{iii}}/{\rm H}_\beta$ ratio. A comparison of galaxies in the group using AAT spectra and classification by \citet{2006MNRAS.372..961K} is shown in Figure \ref{spec}.

Given the evidence of dust presence in the \textit{HST} study, we estimate the dust content using the Balmer decrement \citep{berman} observed from the spectrum. The reddening is $E(B-V)=0.12\pm 0.50$ in the case of $I({\rm H}_\alpha)/I({\rm H}_\beta)= 3.1$, which is expected in the case of AGN activity. We therefore restrict our dust models to have reddening values $0.1,0.2,0.3,0.4,0.5$ in the photometric fits.

The photometric best--fitting template has a solar metallicity, a quickly declining log-normal SFH with $t_0=3 \,{\rm Gyr}$ and  $\tau=0.1$. A low reddening $E(B-V) =0.1$ is preferred, and the stellar mass is $(2.95\pm 0.65)\times 10^{10}M_\odot$. The inclusion of a late SFH burst is disfavored by the fitting apart from intermediate apertures. 

Previous work found that the presence of dust lanes may bias the galaxy stellar mass from unresolved galaxy SED fits to lower values (\citealt{sorba}). The total stellar mass from fitting over the \sextractor segmentation map of NGC4993 is $(3.8\pm 0.20)\times 10^{10} M_\odot$, more than $1\sigma$ higher than the unresolved SED fitting. The specific SFR (sSFR) map from our pixel SED fits is shown in Figure \ref{smass_map}, where the shell structure is clearly visible, suggesting that the sSFR is slightly more accentuated in the stellar halo compared to the inner parts. Younger ages (by $\sim 2~{\rm Gyr}$) are also preferred in the outer regions, though we still do not find evidence for a star formation burst at late times, and explain our results by the stripping of stellar populations from the lower-mass galaxy in a minor dry merger. A dust model with $E(B-V)=0.1$ is preferred in the inner few kpc, while $E(B-V)=0$ is found outside. Despite the presence of dust lanes, an analysis of the \textit{HST} photometry and a comparison with extinction models suggests that the effect of dust is not extreme, with reddening values that are consistent with 0.1 in the core. We therefore believe that the dust obscuration does not play a significant role in our SFR estimates.

\subsection{Pixel Color Diagrams}
In Figure \ref{smass_map} we show a color-magnitude diagram for all the pixels within the field of view of the DECam data near the galaxy. The image has been cleaned of stars and other contamination, thus all points come from the galaxy itself. The position of the GW source, $10.6''$ offset from the center, is the cyan colored pixels, while the center of the galaxy is shown in the red points. This galaxy is well represented by a pixel ``main sequence'' that is bluer at fainter levels, which is typical of early--type galaxy color gradients (e.g., \citealt{lanyon-foster}). We conclude that there is no significant difference between the transient position and other outer light, although it is bluer than the core region. This further supports the scenario in which the BNS formation is not related to some particular recent star formation event in this region.


\section{Discussion and conclusions}\label{implications}

\subsection{BNS formation and delay time under the hypothesis of galaxy merger}

In the most accepted shell formation scenarios the shells are stellar debris coming from the less massive, stripped galaxy, and the arcs form at the apocenter of the orbits of the infalling material \citep{quinn}.

Based on our results, we believe that NGC4993 experienced a dry minor galaxy merger with still visible signs. The shells are expected to be washed out within a time that depends on the velocity dispersion at their position. We estimate the shell survival time in two ways, based on the velocity dispersion of the galaxy as well as the velocity dispersion of the shell itself. From the 6dF spectrum the  line--of--sight central velocity dispersion is $\sigma_v=(160.0\pm 9.1)~{\rm km\, s^{-1}}$. We estimate its value at the position of the transient. 
The velocity dispersion of early type galaxies drops from its central peak value at larger radii, and observations show that the maximum drop to the outer parts of ellipticals near the effective radius is $\sim 40\%$ of the central value (\citealt{emsellem}). Based on the distance of the shell from the center, $R \approx  4 ~{\rm kpc}$, we estimate that the dynamical time at this radius is
$t_{\rm dyn} \equiv  R/\sigma_{v} \approx 60 ~{\rm Myr}$ (the line-of-sight velocity is relevant here, given the shell's geometry, but e.g. if we assume a 3D isotropic velocity dispersion it would reduce the dynamical time by $\sqrt{3}$).

So far we have no measurement of the shell's velocity dispersion, but estimates from the literature suggest for similar shells in other galaxies
$\sigma_v \approx  20 ~{\rm km\,s}^{-1}$ \citep{quinn}. This would give a dynamical time scale of $t_{\rm dyn} \approx  192~ {\rm Myr}$. 
Detailed simulations of  shells in other galaxies suggest that survival time could be even larger that 1 Gyr,  depending on the assumed scenario \citep{pop}.

The survival time of the shell could be used as an upper limit for the time the minor merger took place, i.e  $t_{\rm dyn} \geq t_{\rm merg} $, so we estimate $t_{\rm merg}  \lesssim  200$ Myr.

If the BNS was formed as such in a shell, then we would have expected to see evidence for recent star formation, but we find no indication of this. In the absence of star formation it is plausible that the BNS coalescence was triggered by a dynamical process, e.g. NS-NS capture or the destabilization of a pre-existing wide-separation binary. These processes will be quite sensitive to the stellar density which, given the S\'ersic index and the luminosity from the residual image found in section \ref{morphresults}, is high in the center of NGC4993 and around the transient position. If this dynamical hypothesis is true, then the delay time $\Delta t_{NSM}$ between the BNS formation and coalescence is $\lesssim 200~{\rm Myr}$. On the other hand, \citet{blanchard} find a median delay time of $11^{+0.7}_{-1.4} {\rm Gyr}$ under the assumption that the binary was formed through secular SF.

\subsection{Galaxy environment}
If the binary formation is related to dynamical processes in galaxy merging as we are investigating here, then this is most likely to happen in galaxy groups and low mass clusters.
According to the 2MASS catalog \citep{tully}, NGC4993 resides in a group, of which we analyze the remaining 7 galaxies. A spectral analysis shows that NGC4993 is not the only galaxy showing AGN activity (see Figure \ref{spec}), but it is peculiar in terms of age, metallicity and mass-to-light ratio. It shows an older stellar population (the mean age of the other 13 galaxies is ${\rm Log} (Age) = 9.56\pm0.17$), lower metallicity (mean: $M/H=-0.31 \pm 0.11$), and higher $M/L_r$ (mean: $2.41\pm0.45$ ) than the average. The group has a projected virial radius of $R_{vir} = 0.36$ Mpc and a line-of-sight velocity dispersion $\sigma_v = 143 {\rm km\, s^{-1}}$ (\citealt{tully}). The crossing time is therefore $t_{\rm cr} \sim R_v/(\sqrt{2.5}\sigma_v)\sim 1.6$ Gyr.  

If galaxy mergers are correlated to BNS coalescence, future GW studies could possibly concentrate on galaxy groups (but note that these are crowded regions and therefore matching candidates to a host could be difficult). In order to have precise measurements of $H_0$, one needs to identify the host galaxy redshift clearly. When the match is clear, the properties of the type of host galaxy found could help future studies to select the right host galaxy or create galaxy catalogs of likely hosts for GW EM follow-up and untriggered kilonova searches (\citealt{doctor}). In fact, large photometric surveys such as DES, LSST or WFIRST are expected to observe kilonova events at redshifts beyond the sensitivity of GW experiments, where the angular separation between galaxies decreases (\citealt{scolnic}).

\subsection{BNS merging constraints}
We derive a constraint on neutron stars merging rate at time $t$ by using:
\begin{equation}
R_{NSM}(t) = \alpha R_{NS}(t')\, ,\label{rate}
\end{equation}
where $\alpha$ is the fraction of neutron stars which are in binaries, $t' = t-\Delta t_{NSM}$ and the fraction of mass of formed stars that are NS is: 
\begin{equation}
R_{NS}(t') = \int dM_\star \Phi(M_\star)\Psi(t_\star)\Theta_{NS}(M_\star)\, ,
\end{equation}
with $\Phi(M_\star)$ being the IMF, $\Psi(t_\star)$ is our best fit SFH, $\Theta_{NS}(M_\star)$ is 1 for star mass ranges of $8\, M_\odot<M<20 \,M_\odot$, zero otherwise. We drop the metallicity dependence in $\Theta_{NS}$ because we only consider a solar metallicity for the galaxy, as a result of our spectroscopic fit. $t_\star$ is the time when the progenitor of the NS was formed, therefore satisfying $t' = t_\star+t_{\rm life}$, with $t_{\rm life}$ being the lifetime of the progenitor before becoming a NS. We assume a $t_{\rm life}=0.02$ Gyr, but our calculation is insensitive to this choice as the typical lifetime of these massive stars ($\sim 0.01-0.03$ Gyr) is much shorter than the timescale over which the SFH found for NGC4993 is changing at late times.
We assume a Chabrier IMF, but this choice is not relevant as  we are only exploring the high mass end of the IMF. Assuming $\alpha=0.002$ and the distribution of $\Delta t_{NSM}$ from \citet{vangioni} (their Figure 3 for solar metallicity), and our best fit SFH from Eq. \ref{sfr} with $t_0=3 {\rm Gyr}$ and $\tau=0.3$, we get a NS formation rate of $R_{NS}^{\rm gal}= 3.6^{+28}_{-3.6} \times 10^{-5} ~{\rm yr}^{-1}$ and a BNS merger rate of $R_{NSM}^{\rm gal}= 5.7^{+0.57}_{-3.3} \times 10^{-6} ~{\rm yr}^{-1}$ for the whole galaxy. Errors reflect the uncertainty on the SFH, which dominates our errors: they represent the two central quartiles of the rates distribution computed with the SFHs of the pixel SED fitting over the galaxy.

Given the sensitivity of the BNS merger event rate to the recent SFR of a galaxy, it is somewhat surprising that GW170817 occurred in an old, early type galaxy. We therefore ask what is the probability of observing such an event in any early--type galaxy within the LIGO--detectable volume. To make this estimate we integrate the stellar mass function of early--type galaxies from \citet{weigel} and scale the per--solar--mass rate from Eq. \ref{rate} to the mass contained within the LIGO detectable volume (radius $80~{\rm Mpc}$). We find $R_{NSM}^{\rm early}= 23^{+2}_{-14} ~{\rm yr}^{-1} {\rm Gpc}^{-3}$ resulting in $0.038^{+0.004}_{-0.022}$ expected events. This calculation assumes that the SFH of NGC4993 is representative of local early--type galaxies. In fact much of the mass will be contained in more massive, and on average older and less star--forming, galaxies. We contrast this with a similar calculation for all galaxy types, using the cosmic SFR density from \citet{lognorm}, finding $R_{NSM}^{\rm all}\approx 270 ~{\rm yr}^{-1} {\rm Gpc}^{-3}$ and $\sim 0.5$ expected events.

This result shows that it is unlikely that we observed one such BNS merger with LIGO over the combined nine months of operations in an early--type galaxy. The assumptions in the calculation include the fraction of NS that form in binaries ($\alpha=0.002$) and the delay time distribution, both coming from binary star models (where the progenitors of the BNS were already a bound system) and satisfying Milky Way constraints. If the BNS formation mechanism is via dynamical interaction, our result could point to a higher value of $\alpha$ or a shorter $\Delta t_{NSM}$ for systems that recently underwent a galaxy merger, more so for those that have high stellar density (such as early--type galaxies). It is therefore of interest to know the fraction of galaxies similar to NGC4993 that show similar signs of a galaxy merger in the form of visible shells. We select galaxies from the first year of DES data with size, surface brightness and S\'ersic index within $10\%$ of the best fit values for NGC4993. We find $1100$ such galaxies, and visually inspect them to identify shell galaxies. Only $15\%$ of these objects display shells, and so NGC4993 is unusual amongst early-type galaxies. This is far from conclusive evidence for a merger origin of BNS events. However, the coincidence of evidence for a recent merger in a galaxy for which a BNS event was otherwise improbable is compelling.

\acknowledgments
Funding for the DES Projects has been provided by the DOE and NSF(USA), MEC/MICINN/MINECO(Spain), STFC(UK), HEFCE(UK). NCSA(UIUC), KICP(U. Chicago), CCAPP(Ohio State), 
MIFPA(Texas A\&M), CNPQ, FAPERJ, FINEP (Brazil), DFG(Germany) and the Collaborating Institutions in the Dark Energy Survey.

The Collaborating Institutions are Argonne Lab, UC Santa Cruz, University of Cambridge, CIEMAT-Madrid, University of Chicago, University College London, 
DES-Brazil Consortium, University of Edinburgh, ETH Z{\"u}rich, Fermilab, University of Illinois, ICE (IEEC-CSIC), IFAE Barcelona, Lawrence Berkeley Lab, 
LMU M{\"u}nchen and the associated Excellence Cluster Universe, University of Michigan, NOAO, University of Nottingham, Ohio State University, University of 
Pennsylvania, University of Portsmouth, SLAC National Lab, Stanford University, University of Sussex, Texas A\&M University, and the OzDES Membership Consortium.

Based in part on observations at Cerro Tololo Inter-American Observatory, National Optical Astronomy Observatory, which is operated by the Association of 
Universities for Research in Astronomy (AURA) under a cooperative agreement with the National Science Foundation.

The DES Data Management System is supported by the NSF under Grant Numbers AST-1138766 and AST-1536171. 
The DES participants from Spanish institutions are partially supported by MINECO under grants AYA2015-71825, ESP2015-66861, FPA2015-68048, SEV-2016-0588, SEV-2016-0597, and MDM-2015-0509, 
some of which include ERDF funds from the European Union. IFAE is partially funded by the CERCA program of the Generalitat de Catalunya.
Research leading to these results has received funding from the European Research
Council under the European Union's Seventh Framework Program (FP7/2007-2013) including ERC grant agreements 240672, 291329, and 306478.
We  acknowledge support from the Australian Research Council Centre of Excellence for All-sky Astrophysics (CAASTRO), through project number CE110001020.

This manuscript has been authored by Fermi Research Alliance, LLC under Contract No. DE-AC02-07CH11359 with the U.S. Department of Energy, Office of Science, Office of High Energy Physics. The United States Government retains and the publisher, by accepting the article for publication, acknowledges that the United States Government retains a non-exclusive, paid-up, irrevocable, world-wide license to publish or reproduce the published form of this manuscript, or allow others to do so, for United States Government purposes.

AP, WH and OL are supported by ERC Advanced Grant FP7/291329.\\
Based in part on data acquired through the Australian Astronomical Observatory. \\
Based on data obtained from the Mikulski Archive for Space Telescopes (MAST). STScI is operated by the Association of Universities for Research in Astronomy, Inc., under NASA contract NAS5-26555. Support for MAST for non-HST data is provided by the NASA Office of Space Science via grant NNX09AF08G and by other grants and contracts.

\bibliographystyle{yahapj}

\end{document}